\documentclass[english,aps,pra,superscriptaddress,showpacs,twocolumn,floatfix,tocloft]{revtex4}

\usepackage{amsmath}
\usepackage{amssymb}
\usepackage{amsfonts}
\usepackage{bbm}
\usepackage{epsfig}
\usepackage{grffile}
\usepackage{babel}

\usepackage[usenames,dvipsnames]{color}
\definecolor{grey}{rgb}{.5,.5,.5}
\definecolor{dblue}{rgb}{0,0,.5}
\definecolor{dgreen}{rgb}{0,.65,0}

\usepackage[colorlinks=true,citecolor=dblue,linkcolor=dblue]{hyperref}
\usepackage[all]{hypcap}
\usepackage{setspace}
\DeclareRobustCommand{\SkipTocEntry}[5]{}

\newcommand{\id}{\mathbbm{1}}
\newcommand{\Tr}{\operatorname{Tr}}
\newcommand{\aux}{{\mathrm{aux}}}
\newcommand{\odd}{{\mathrm{odd}}}
\newcommand{\even}{{\mathrm{even}}}
\newcommand{\bra}{\langle}
\newcommand{\ket}{\rangle}
\newcommand{\mc}[1]{\mathcal{#1}}
\newcommand{\pdag}{{\phantom{\dag}}}

\newcommand{\hH}{\hat{H}}
\newcommand{\hS}{\hat{S}}
\newcommand{\hX}{\hat{X}}
\newcommand{\hY}{\hat{Y}}
\newcommand{\dm}{{\hat{\rho}}}
\newcommand{\dmp}{\rho}
\newcommand{\epsP}{\epsilon_{\mathrm{P}}}
\newcommand{\epsM}{\epsilon_{\mathrm{M}}}
\newcommand{\gs}{{\mathrm{gs}}}
\newcommand{\mre}{\mathrm{e}}
\newcommand{\mri}{\mathrm{i}}
\renewcommand{\vec}[1]{{\boldsymbol{#1}}}

\newcommand{\normS}[1]{\Vert #1\Vert}

\newcommand{\vn}{{\vec{n}}}
\newcommand{\vs}{{\vec{\sigma}}}

\usepackage{array}
\usepackage{multirow}
\usepackage{hhline}
\newcommand\T{\rule{0pt}{2.6ex}}       
\newcommand\TT{\rule{0pt}{4ex}}       
\newcommand\B{\rule[-1.2ex]{0pt}{0pt}} 
\newcommand\BB{\rule[-2ex]{0pt}{0pt}} 
\newcommand{\SpaceCaptionTable}{\vspace{0.3cm}}

\newcommand{\lmu} {Department of Physics,
Ludwig-Maximilians-Universit{\"a}t M{\"u}nchen, Theresienstr.\ 37, 80333 Munich, Germany}
\newcommand{\psud} {Laboratoire de Physique Th\'{e}orique et Mod\`{e}les Statistiques, Universit\'{e} Paris-Sud, CNRS UMR 8626, 91405 Orsay Cedex, France}
\newcommand{\duke} {Department of Physics, Duke University, Durham, North Carolina 27708, USA}

\begin{document}

\title{Minimally entangled typical thermal states versus matrix product purifications\\ for the simulation of equilibrium states and time evolution}
\author{Moritz Binder}
\affiliation{\duke}
\affiliation{\lmu}
\author{Thomas Barthel}
\affiliation{\duke}
\affiliation{\psud}
\date{November 11, 2014}

\begin{abstract}
For the simulation of equilibrium states and finite-temperature response functions of strongly-correlated quantum many-body systems, we compare the efficiencies of two different approaches in the framework of the density matrix renormalization group (DMRG). The first is based on matrix product purifications. The second, more recent one, is based on so-called minimally entangled typical thermal states (METTS). For the latter, we highlight the interplay of statistical and DMRG truncation errors, discuss the use of self-averaging effects, and describe schemes for the computation of response functions. For critical as well as gapped phases of the spin-$1/2$ XXZ chain and the one-dimensional Bose-Hubbard model, we assess the computation costs and accuracies of the two methods at different temperatures. For almost all considered cases, we find that, for the same computation cost, purifications yield more accurate results than METTS -- often by orders of magnitude. The METTS algorithm becomes more efficient only for temperatures well below the system's energy gap. The exponential growth of the computation cost in the evaluation of response functions limits the attainable timescales in both methods and we find that in this regard, METTS do not outperform purifications.
\end{abstract}


\maketitle

\begin{spacing}{0.95}
\tableofcontents
\end{spacing}

\section{Introduction}
Finite-temperature correlation and response functions of quantum many-particle systems are of great interest. They provide insights into the many-body physics and allow to compare theoretical models to experimental results. However, their accurate computation remains challenging. For many relevant models, one has to rely on the development of efficient numerical techniques. The most successful method for the study of strongly correlated one-dimensional (1D) systems is the density-matrix renormalization group (DMRG), which is based on matrix product states (MPS) \cite{White1992-11,White1993-10,Schollwoeck2005}. While DMRG was originally designed to study ground states of 1D systems, its extension to the time evolution of quantum states within tDMRG  \cite{Vidal2003-10,White2004,Daley2004} allows for the simulation of quenches and response functions. Based on this extension, quite different methods for simulations at finite temperatures have been developed. One of them rests on a purification \footnote{A state $|\mathrm{P}_\dm\ket\in\mc{H}\otimes\mc{H}_\aux$ is called a purification of the density matrix $\dm$ on $\mc{H}$ if $\Tr_\aux |\mathrm{P}_\dm\ket\bra \mathrm{P}_\dm|=\dm$.} of the density matrix \cite{Uhlmann1976,Uhlmann1986,Nielsen2000}, which can be encoded in matrix product form \cite{Verstraete2004-6,Zwolak2004-93}. First, this was successfully applied to study static finite-temperature properties of quantum spin chains \cite{Feiguin2005-72,Barthel2005}. The combination with real-time tDMRG allows for the accurate evaluation of finite-temperature response functions and can be applied to compute spectral functions and to study a variety of experimentally relevant systems \cite{Barthel2009-79b,Feiguin2010-81,Karrasch2012-108,Barthel2012_12,Barthel2013-15,Karrasch2013-15,Karrasch2013-87,Lake2013-111,Huang2013-88}. For such purposes, purifications can similarly be combined with a Chebyshev expansion \cite{Tiegel2014-90}. Despite their success, simulations based on purifications are often limited with respect to the reachable inverse temperatures, times, or Krylov expansion orders in frequency domain approaches. This is due to a growth of entanglement, which is accompanied by a corresponding growth of computation costs. The search for complementary approaches led to an algorithm that avoids the direct encoding of the mixed states: Instead of purifying the density matrix, one can sample cleverly chosen pure states, so-called minimally entangled typical thermal states (METTS) \cite{White2009-102,Stoudenmire2010-12}. While they can be efficiently encoded in matrix product form as their entanglement is relatively low, they represent well the thermal properties of the system at hand. The METTS algorithm has been successfully applied to study static properties and quantum quenches at finite temperature \cite{Yao2012,Alvarez2013-87,Bonnes2014-113}. However, a thorough analysis of its accuracy and efficiency compared to computations using matrix product purifications of the density matrices was lacking.

The extreme cases of infinite and zero temperatures are relatively easy to understand. Purifications should prevail at higher temperatures, while METTS should become favorable at lower temperatures when the ground state is approached. The infinite temperature purification can be written exactly as an MPS of bond dimension $D=1$. This is clearly simpler and more efficient than averaging over several METTS samples (each having in this case the same probability). The argument prevails for large finite temperatures. As $\mre^{-\beta\hH} \approx 1-\beta\hH$ can for example be represented exactly as a purification of bond dimension $D=5$ for the XXZ model, the evaluation of an arbitrary product operator in a chain of length $L$ would cost $\mc O(LD^3)$ operations. In contrast, the number of METTS samples required for a precise estimate should scale exponentially with the size of the spatial support of the considered product operator. On the other hand, at zero temperature and for a system without ground-state degeneracy, (almost) every METTS is simply the ground state $|\gs\ket$. So, one only needs to produce a single METTS with bond dimension $D_\gs$. In contrast, the zero-temperature purification would correspond to the projector $|\gs\ket\bra\gs|$ and have bond dimension $D_\gs^2$. Thus, for $T=0$, METTS are clearly more efficient.

In this work, we discuss how self-averaging can be used to moderately reduce statistical errors in the METTS algorithm and introduce schemes for the evaluation of response functions using METTS. We compare the accuracies and computation costs of the METTS and purification approaches for the evaluation of finite-temperature correlation and response functions. We focus on two paradigmatic models of interacting quantum systems, namely the spin-$1/2$ XXZ chain \cite{Bethe1931,Cloizeaux1966-7,Mikeska2004} and the 1D Bose-Hubbard model \cite{Kuehner1998-58,Jaksch1998-81} at critical as well as non-critical points of their phase diagrams. In contrast to indications and expectations expressed in the earlier literature \cite{White2009-102,Schollwoeck2009-2,Bonnes2014-113}, for almost all cases considered here, we find that, for the same computation cost, the purification approach yields more accurate results than METTS -- often by orders of magnitude. METTS become more efficient only for temperatures well below the energy gap of the system. It would be interesting to investigate further whether other approaches as, for example, computations based on the ground state and a few excited states, which can be determined variationally, could outperform the METTS approach for such very low temperatures. For the comparisons, we always use equal total computation costs for both methods, ignoring that METTS simulations can be parallelized more easily than purification simulations by generating independent Markov chains on different computing nodes. This can be taken into account by keeping in mind that, to reduce the presented METTS errors by an order of magnitude, one needs to increase the number of employed computing nodes by at least a factor of 100.

The article is structured as follows. Sections~\ref{Sec:Purification} and \ref{Sec:METTS} shortly review the algorithms for computing static observables with purifications and METTS, respectively, and discuss the interplay of statistical and truncation errors for METTS (Sec.~\ref{Sec:METTS_StatisticsVsTrunc}) as well as self-averaging (Sec.~\ref{Sec:METTS_SelfAverage}). In section~\ref{Sec:Response}, we introduce, for METTS, a simple scheme and two more elaborate schemes for the computation of response functions, which are analogous to corresponding schemes based on purifications \cite{Barthel2009-79b,Karrasch2012-108,Barthel2012_12}. The main objective of the paper, the efficiency comparison of METTS and purifications, is presented in section~\ref{Sec:Statics} for static correlations functions in the spin-$1/2$ XXZ chain and the Bose-Hubbard model, and in section~\ref{Sec:ResponseXXZ}, for response functions in the XXZ model. Some technical issues are described in appendices. We summarize and conclude in section~\ref{Sec:Conclusion}.

\section{Matrix product purifications} \label{Sec:Purification}
Let us briefly review how to compute finite-temperature expectation values $\bra\hat{O}\ket_\beta=\Tr(\dm_\beta\hat{O})/Z_\beta$ using matrix product purifications. Here, we work with the canonical ensemble $\dm_\beta=\exp(-\beta\hH)$ and $Z_\beta=\Tr\dm_\beta$.

A state $|\mathrm{P}_\dm\ket\in\mc{H}\otimes\mc{H}_\aux$ is called a purification of the density matrix $\dm$ on $\mc{H}$ if
\begin{equation}\label{eq:purify}
	\Tr_\aux |\mathrm{P}_\dm\ket\bra \mathrm{P}_\dm|=\dm.
\end{equation}
Choosing the auxiliary Hilbert space $\mc{H}_\aux$ isomorphic to the physical Hilbert space $\mc{H}$, i.e., $\mc{H}\simeq\mc{H}_\aux$, it is simple to give a purification of the infinite-temperature state $\dm_0=\id$. It is
\begin{equation}\label{eq:dm0}
	|\mathrm{P}_{\dm_0}\ket = \bigotimes_i \big( \sum_{\sigma_{i}}|\sigma_{i}\ket \otimes |\sigma_{i}\ket_\aux \big),
\end{equation}
where $|\sigma_{i}\ket$ are orthonormal basis states for lattice site $i$, and $|\sigma_{i}\ket_\aux$ for the corresponding lattice site of the auxiliary system.
For the orthonormal basis $\{|\vs\ket=\bigotimes_i|\sigma_i\ket\}$ of $\mc{H}$, let $|X\ket\in\mc{H}\otimes \mc{H}_\aux$ denote the vectorization of an operator $\hat{X}$ on $\mc{H}$ such that
\begin{equation}\label{eq:vectorize}
	|X\ket \equiv \sum_{\vs,\vs'}\bra\vs|\hat{X}|\vs'\ket \,|\vs\ket\otimes|\vs'\ket_\aux.
\end{equation}
In this notation, we have that $|\dmp_{\beta/2}\ket\in \mc{H}\otimes\mc{H}_\aux$ is according to equations \eqref{eq:purify} and \eqref{eq:vectorize} a purification of the density matrix $\dm_\beta$.

Because $|\dmp_0\ket\equiv |\mathrm{P}_{\dm_0}\ket$, as given in Eq.~\eqref{eq:dm0}, is a product state, it can be encoded as an MPS with matrices of bond dimension one (cf.\ appendix~\ref{Appx:Truncations}). With $|\dmp_0\ket$ as the initial state, one can employ imaginary-time evolution, to obtain purifications $|\dmp_{\beta/2}\ket$ for finite-temperature states $\dm_\beta$,
\begin{equation}\label{eq:purify2}
	|\dmp_{\beta/2}\ket = \big(\mre^{-\beta\hH/2} \otimes \id_\aux\big) |\dmp_0\ket.
\end{equation}
To this purpose, one can employ the time-dependent DMRG algorithm (tDMRG) \cite{White2004,Daley2004} or the almost identical time-evolved block decimation (TEBD) \cite{Vidal2003-10}. Specifics of our simulations are summarized in appendix~\ref{Appx:Truncations}. Exploiting that
\begin{equation*}
	Z_\beta=\bra\dmp_{\beta/2}|\dmp_{\beta/2}\ket\quad \text{and}\quad
	\dm_\beta =\Tr_\aux |\dmp_{\beta/2}\ket \bra\dmp_{\beta/2}|,
\end{equation*}
thermal expectation values can be computed by
\begin{equation}
	\bra \hat{O} \ket_\beta = \frac{1}{Z_\beta} \Tr\big(\mre^{-\beta\hH} \hat{O}\big) = \frac{\bra\dmp_{\beta/2}|\hat{O}|\dmp_{\beta/2}\ket}{\bra\dmp_{\beta/2}|\dmp_{\beta/2}\ket},
\end{equation}
where both physical and auxiliary degrees of freedom are summed over.

\section{METTS sampling} \label{Sec:METTS}
\subsection{Algorithm for static observables} \label{Sec:METTSalgo}
\begin{figure}
\label{fig:SchematicMETTSSampling}
\includegraphics[width = \columnwidth]{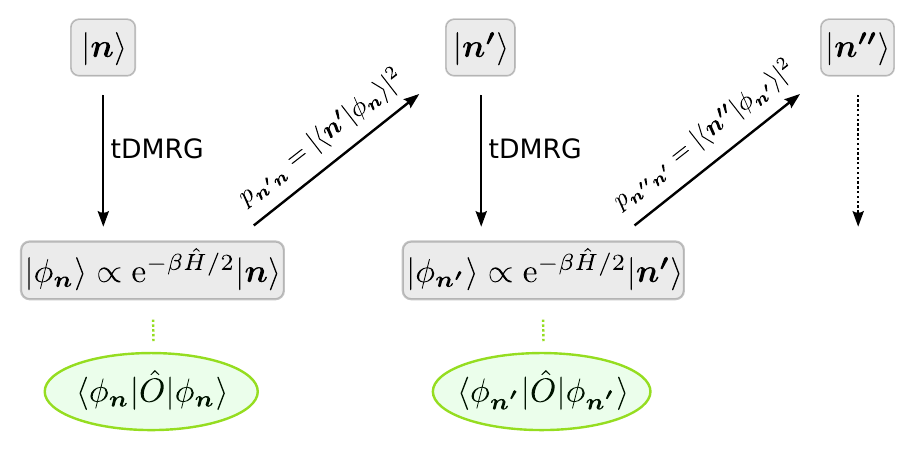}
\caption{METTS algorithm for the evaluation of static observables. A product state $|\vn\ket$ is evolved in imaginary time up to $\tau=\beta/2$ and normalized to obtain the METTS sample $|\phi_{\vn}\ket$. A projective measurement with transition probabilities $p_{\vn'\vn}=|\bra\vn'|\phi_\vn\ket|^2$ yields a new product state $|\vn'\ket$, which is again evolved in imaginary time, etc. Observables are evaluated by averaging the expectation values obtained from the samples.}
\end{figure}

The strategy employed in the minimally entangled typical thermal states (METTS) algorithm is to approximate thermal expectation values $\bra \hat{O} \ket_\beta$ by sampling pure quantum states that have two favorable properties. They represent well the physical properties of the system for the given temperature, and the entanglement of the states is relatively low, which makes DMRG calculations efficient.
Expressing the trace for the thermal expectation value using some orthonormal basis $\{|\vn\ket\}$ of product states
\begin{equation}\label{eq:METTSbasis}
	|\vn\ket = \bigotimes_i|n_i\ket,
\end{equation}
where $|n_i\ket$ are (arbitrary) orthonormal basis states for lattice site $i$, we have
\begin{equation}\label{eq:METTStrace}
	\bra \hat{O} \ket_\beta = \frac{1}{Z_\beta} \sum_\vn \bra \vn|\mre^{-\beta \hH} \hat{O} |\vn\ket.
\end{equation}
Defining the METTS $|\phi_\vn\ket$ and their probabilities $P_\vn$, 
\begin{equation}
	|\phi_\vn\ket := \frac{1}{\sqrt{P_\vn}}\mre^{-\beta \hH/2}|\vn\ket,\quad
	P_\vn:= \bra \vn | \mre^{-\beta \hH} | \vn\ket,
\end{equation}
the thermal average reads
\begin{equation}\label{eq:METTS_expectation}
	\bra \hat{O} \ket_\beta = \frac{1}{Z_\beta} \sum_\vn P_\vn \bra \phi_\vn|\hat{O}|\phi_\vn\ket.
\end{equation}
Thus, by sampling the states $|\phi_\vn\ket$ according to the probabilities $P_\vn/Z_\beta$, we can approximate $\bra \hat{O} \ket_\beta$ by averaging over $\bra \phi_\vn |\hat{O}|\phi_\vn\ket$. The computation cost of DMRG is directly related to the entanglement of the quantum state (see appendix~\ref{Appx:Computation_cost}). Hence, product states \eqref{eq:METTSbasis} are a natural choice because their entanglement entropy is zero and it remains reasonably low during imaginary-time evolution. The sampling is accomplished efficiently by generating a Markov chain of METTS as illustrated in Fig.~\ref{fig:SchematicMETTSSampling}. An arbitrary initial product state $|\vn\ket$ is evolved in imaginary time to obtain the METTS $|\phi_\vn\ket$. (See appendix~\ref{Appx:Truncations} for details on the tDMRG evolution.) Then, the METTS is collapsed through a projective measurement with measurement basis \eqref{eq:METTSbasis}, yielding a new product state $|\vn'\ket$ with probability $p_{\vn'\vn}:=|\bra\vn'|\phi_\vn\ket|^2$ from which one subsequently computes $|\phi_{\vn'}\ket$ and so on. The transition probabilities obey the detailed balance condition
\begin{equation}
	p_{\vn'\vn}P_\vn=|\bra\vn'|\mre^{-\beta\hH/2}|\vn\ket|^2=p_{\vn\vn'}P_{\vn'}
\end{equation}
such that the desired distribution $P_\vn$ is indeed the fixed point of this Markov process.

Note that the projective measurement, $|\phi_\vn\ket\to |\vn'\ket$, can be carried out sequentially, site by site. Starting at some site $i$, we go from $|\phi_\vn\ket$ to $|n'_i\ket\bra n'_i|\allowbreak\cdot|\phi_\vn\ket/\sqrt{\pi(n'_i)}$ with probability $\pi(n'_i):=|{\bra n'_i|\phi_\vn\ket}|^2$. Measuring subsequently on site $j$, we go to $|n'_i n'_j\ket\bra n'_i n'_j|\cdot|\phi_\vn\ket/\sqrt{\pi(n'_i)\pi(n'_j|n'_i)}$ with cond.\ probability $\pi(n'_j|n'_i):=|{\bra n'_i n'_j|\phi_\vn\ket}|^2/\pi(n'_i)$ such that, in the end, we arrive at state $|\vn'\ket$ indeed with probability 
\begin{equation*}
	p_{\vn'\vn}=\pi(n'_i)\pi(n'_j|n'_i)\pi(n'_k|n'_in'_j)\dotsc =|\bra\vn'|\phi_\vn\ket|^2.
\end{equation*}
Due to this, the projective measurement of MPS $|\phi_\vn\ket$ can be done efficiently in a single sweep through the lattice \cite{White2009-102,Stoudenmire2010-12}. 

In order to ensure ergodicity and reduce autocorrelation times, it is useful to switch between different measurement bases $\{|\vn^{(k)}\ket\}$ during the sampling. Details on this and our corresponding choice are described in appendix~\ref{Appx:METTS_collapse}.

\subsection{Errors: Statistics, truncations, and Trotter} \label{Sec:METTS_StatisticsVsTrunc}
\begin{figure}[t]
\label{fig:convergence}
\includegraphics[width=\columnwidth]{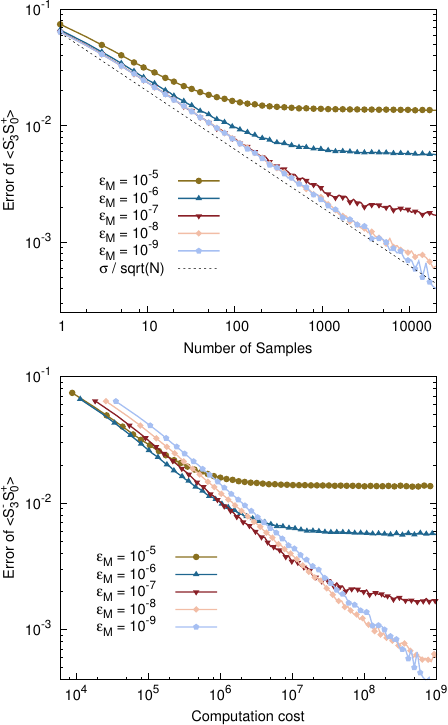}
\caption{Convergence of the METTS algorithm for the static correlation function $\bra \hS^{-}_{3}\hS^{+}_{0} \ket_\beta$ in an XX chain [$\Delta=0$ in Eq.~\eqref{eq:H_XXZ}] of length $L=64$ at $\beta = 4$. The figure shows the errors for different truncation thresholds $\epsM$ as a function of the number of samples (top) and as a function of the computation cost (bottom).}
\end{figure}
There are two error sources for the evolution of MPS in the framework of tDMRG as described in appendix~\ref{Appx:Truncations}. The first is due to truncations of low-weight terms in the Schmidt decomposition of the wave function. This error is well-controlled by the truncation threshold (we call it $\epsP$ for purifications and $\epsM$ for METTS) which bounds, in every time step, the two-norm deviation $\normS{\psi_{\operatorname{trunc}} - \psi}$ of the truncated MPS from the exactly evolved state. We implement tDMRG using fourth-order Trotter-Suzuki decompositions of the evolution operators with time steps $\Delta t$ of size $0.125$ for purifications and $0.05$ for METTS ($\hbar=1$). These are the second error source. The resulting errors of order $\Delta t^5$ can only become relevant for very large times and we made sure that they are never dominant for the presented data. Additionally, the accuracy of the METTS sampling algorithm is influenced by a third error source -- the statistical error that depends on the number of samples $N$ used for averaging. 

Figure~\ref{fig:convergence} illustrates the interplay of statistical errors and truncation errors in a METTS computation of the correlator $\bra S^{-}_{3} S^{+}_{0}\ket_\beta$ for the exactly solvable 1D XX model [$\Delta=0$ in Eq.~\eqref{eq:H_XXZ}] at inverse temperature $\beta=4$, where site $i=0$ is at the center of the chain. The exact solution is shortly described in appendix~\ref{Appx:ReferenceData}. The top panel shows the convergence of the METTS algorithm for different truncation thresholds $\epsM$ as a function of the number of samples. For low sample numbers, the statistical error dominates and the truncation error is negligible. Autocorrelation times between subsequent samples are short and the statistical error is to a good approximation proportional to $1/\sqrt{N}$ and independent of the truncation threshold.
Once the sample number reaches a certain $\epsM$-dependent threshold, the statistical error has reduced to a magnitude that is comparable to the error induced by the truncations. The curves begin to level out as the relative contribution of the truncation error grows. At a certain point, further samples will not enhance the accuracy of the simulation as the truncations prevent further convergence to the correct value. Of course, the less we truncate (the lower $\epsM$ is), the longer the $1/\sqrt{N}$-convergence persists.

The truncation affects the accuracy in two ways. As every METTS is approximated by a truncated MPS, the values one obtains for each sample are not exact. Additionally, the produced samples will not correspond exactly to the correct probability distribution $P_n$ of Eq.~\eqref{eq:METTS_expectation}. This is because the transition probabilities depend on the samples and are thus also affected by the truncations.

While lowering the truncation threshold $\epsM$ yields more accurate results, it also increases the computation cost per sample. In the lower panel of Fig.~\ref{fig:convergence}, we present the same errors of the METTS algorithm as in the top panel, but here as a function of the total computation cost which we quantify in an implementation-independent way as described in appendix~\ref{Appx:Computation_cost}. This shifts the simulations with lower truncation thresholds and thus more costly samples to the right. Whereas, for a fixed number of samples (top panel), the accuracy is a monotonic function of $\epsM$, this is not necessarily so for fixed total computation cost (lower panel). When plotted against the total computation costs, accuracy curves of METTS simulations with different $\epsM$ have crossings. For practical simulations, it is therefore important to choose the truncation threshold such that the two error sources are balanced. In the lower panel of Fig.~\ref{fig:convergence}, one can easily read off the truncation threshold that is optimal for a given computation cost. While the optimal truncation threshold depends on the specific system studied and the observable that is evaluated, it generally shifts towards lower values of $\epsM$ with increasing total computation cost.

\subsection{Exploiting self-averaging} \label{Sec:METTS_SelfAverage}
\begin{figure}[b]
\label{fig:SelfAveraging}
\includegraphics[width=\columnwidth]{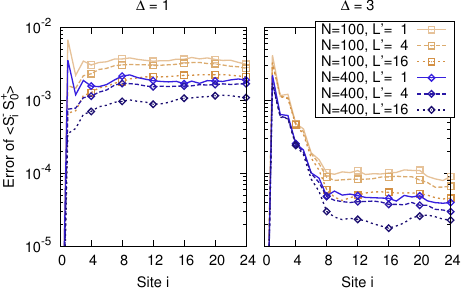}
\caption{Reducing statistical errors in the METTS algorithm by self-averaging. Shown are errors for the  correlator $\bra \hat{S}^{-}_{i} \hat{S}^{+}_{0}\ket_\beta$ in an XXZ chain \eqref{eq:H_XXZ} of length $L=64$ at inverse temperature $\beta = 4$. Fixing the truncation threshold to $\epsilon_{\mathrm{M}} = 10^{-12}$ and the number of samples to $N=100,400$, the correlator was estimated by averaging over $L'=1$, $4$ or $16$ central sites, respectively.}
\end{figure}
If the considered model is translation invariant, we are free to choose an arbitrary position in the lattice for the evaluation of an observable. While the average of these expectation values will converge to the correct result independent of the position, a single METTS sample itself is not translation invariant. We can thus exploit self-averaging to reduce statistical errors in the METTS algorithm. For a correlation length $\xi$, averaging a local observable $\bra\hat{O}_{x_{0}}\ket_\beta$ over a block of $L'$ sites $x_{0}$ corresponds for high and intermediate temperatures to $\mc{O}(L'/\xi)$ statistically independent samples. Therefore, we can expect that the statistical METTS errors reduce by a factor of order $\sqrt{\xi/L'}$. For systems with open boundary conditions, one has to restrict the averaging to sites $x_{0}$ with a sufficient distance from the boundaries.

We illustrate the effect for the spin-$1/2$ XXZ chain in Fig.~\ref{fig:SelfAveraging}, by averaging the correlator $\bra \hat{S}^{-}_{x_0+i} \hat{S}^{+}_{x_0}\ket_\beta$ over different numbers $L'$ of central sites $x_0$ and comparing the result to quasi-exact purification data (see appendix~\ref{Appx:ReferenceData}). The observed error reduction is of the expected order of magnitude. For $\Delta=3$ and $\beta=4$, the impact of self-averaging is rather small for short distances $i$. This is due to the fact that, in this case, the temperature is already well below the gap (cf.\ Table~\ref{tab:gaps}) and all excitations occurring in the METTS are of long wavelength. Hence, short-range correlations in the METTS are almost translation invariant.

As the additional computation cost for the spatial averaging of time-local observables (in equilibrium or quench dynamics) is negligible, it is advisable to enhance the METTS precision through the self-averaging whenever finite-size effects are well-controlled. Spatial averaging in the evaluation of a response function would however require additional real-time evolutions and seems hence not useful.

\section{Comparison for static correlators} \label{Sec:Statics}

\subsection{Procedures to compare efficiencies} \label{Sec:Efficiency}
In this section, we compare the efficiencies of METTS and matrix product purifications, by studying accuracies of static thermal correlation functions for fixed computation costs. To this purpose, accuracies are quantified by the deviations of the obtained expectation values from exact or quasi-exact data as described in appendix~\ref{Appx:ReferenceData}. For the error of $N$ METTS, we generate several sets of $N$ subsequent samples, and take the root mean square of the average deviations from the reference data in each set. The computation cost due to evolving MPS in imaginary-time (and also real-time in section~\ref{Sec:Response}) is quantified in a fashion that is largely independent of the chosen implementation, time step, and platform, as a function of the MPS bond dimensions $D_i=D_i(\beta,t)$ as detailed in appendix~\ref{Appx:Computation_cost}. For METTS, we average the computation costs of the sample sets.
\begin{figure*}[t]
\center
\label{fig:XXZ-static}
\includegraphics[width=\textwidth]{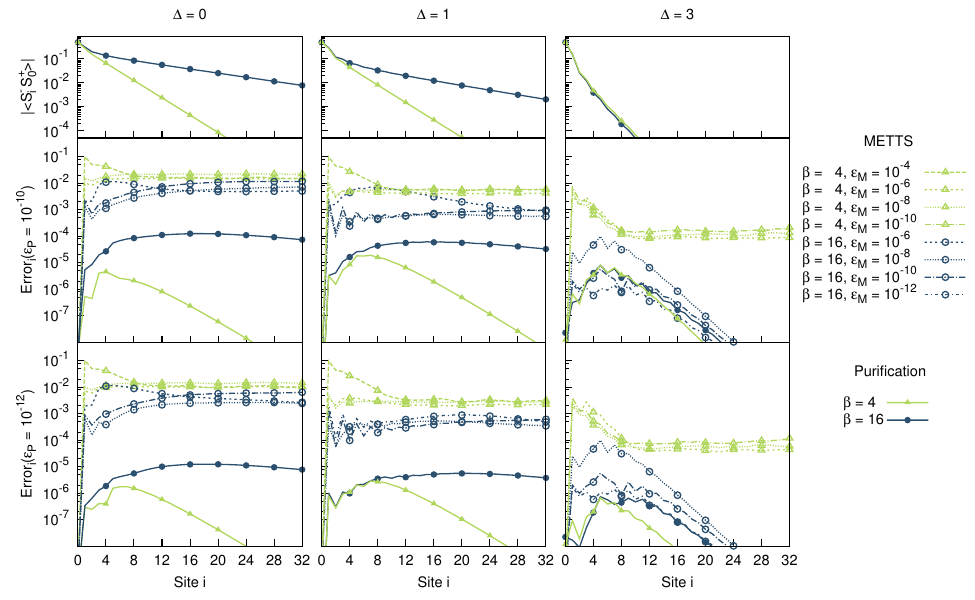}
\caption{Comparing accuracies of METTS and purifications for the static correlation function $\bra \hS^{-}_{i}\hS^{+}_{0} \ket_\beta$ in a spin-$1/2$ XXZ chain \eqref{eq:H_XXZ} of length $L=128$ at inverse temperatures $\beta = 4$, $16$. Site $i=0$ is at the center of the chain. The columns refer to the non-interacting case ($\Delta=0$, left), the isotropic Heisenberg antiferromagnet ($\Delta=1$, center) and the gapped N\'{e}el phase ($\Delta=3$, right). The top panels show the (quasi-)exact expectation values. Below we compare the errors of the purification with truncation thresholds $\epsP = 10^{-10}$, $10^{-12}$ to several METTS simulations with different truncation thresholds $\epsM$. All curves in a given panel that refer to the same temperature share the same computation cost. The corresponding METTS sample set sizes and bond dimensions are specified in Table~\ref{tab:NumberMETTS-XXZ}.}
\end{figure*}

To take the interplay between statistical and truncation errors (section~\ref{Sec:METTS_StatisticsVsTrunc}) into account in the assessment of the performances of METTS and purifications, we proceed as follows. We choose a truncation threshold $\epsP$ for the purification and determine the theoretical computation cost of the simulation. Then we produce METTS samples, using different truncation thresholds $\epsM$, and fix the sample set size $N$ for each $\epsM$ by the quotient of the purification cost and the average cost per METTS sample such that, for each truncation threshold, the total computation costs of both methods are equal. In the figures, we only present the results for the truncation thresholds $\epsM$ that yield the best results, i.e., those $\epsM$ that approximately minimize the error for fixed computation cost, as well as the results for two nearby values of $\epsM$.

Setting the total computation costs of both methods equal, ignores that METTS simulations can be parallelized more easily than purification simulations, by generating independent Markov chains on different computing nodes. However, this can be easily taken into account. To reduce the METTS errors, as presented in the following, by an order of magnitude, one has to increase the number of employed computing nodes by at least a factor of 100. This assumes that the extent of thermalization phases at the beginnings of the Markov chains is negligible, and the factor 100 is a lower bound because one also needs to decrease $\epsM$ (hence, increasing the computation cost per METTS) when the statistical error is being reduced.

\subsection{Spin-\texorpdfstring{$1/2$}{1\textfractionsolidus 2} XXZ chain} \label{Sec:XXZ}
For the spin-$1/2$ XXZ chain \cite{Bethe1931,Cloizeaux1966-7,Mikeska2004} with Hamiltonian
\begin{equation}\label{eq:H_XXZ}
	\hH = \sum_i \left\{ \frac{1}{2} \left( \hS^{+}_{i} \hS^{-}_{i+1} + \hS^{-}_{i} \hS^{+}_{i+1} \right) + \Delta \hS^{z}_{i} \hS^{z}_{i+1} \right\},
\end{equation}
let us consider three values of the anisotropy parameter: the exactly solvable non-interacting case $\Delta=0$ and the isotropic Heisenberg antiferromagnet at $\Delta=1$, which are both critical (gapless), as well as the point $\Delta=3$ in the gapped N\'{e}el phase. The lattice with open boundary conditions has size $L$ and site $i=0$ is at the center of the chain. We apply the METTS and purification algorithms, as described in sections~\ref{Sec:Purification} and \ref{Sec:METTSalgo}, to compute the static correlation function $\bra \hS^{-}_{i} \hS^{+}_{0} \ket_\beta$ at inverse temperatures $\beta=4$ and $16$. 
\begin{table}[b]
\label{tab:gaps}
\begin{spacing}{1.2}
Spin-$1/2$ XXZ chain\\[0.4em]
\begin{tabular}{@{\quad}l@{\quad}|@{\quad}c@{\quad}c@{\quad}c@{\quad}}
                   &$\Delta=0$	&$\Delta=1$	&$\Delta=3$\\ \hline
 $\Delta E_{L=32}$ &$0.0475(8)$	&$0.1177(4)$	&$0.6521(9)$\\
 $\Delta E_{L=64}$ &$0.0241(6)$	&$0.0617(6)$	&$0.6242(5)$\\
 $\Delta E_{L=128}$&$0.0121(8)$	&$0.0319(3)$	&$0.6162(7)$
\end{tabular}\\[1.2em]
1D Bose-Hubbard model\\[0.4em]
\begin{tabular}{@{\quad}l@{\quad}|@{\quad}c@{\quad}c@{\quad}c@{\quad}}
                   &$J=0.09375$	&$J=0.125$	&$J=0.25$\\ \hline
 $\Delta E_{L=32}$ &$0.1415(6)$	&$0.032(6)$	&$0.0134(4)$\\
 $\Delta E_{L=64}$ &$0.1395(2)$	&$0.0291(6)$	&$0.0005(0)$\\
 $\Delta E_{L=128}$&$0.1389(7)$	&$0.0282(0)$	&$0.0001(2)$
\end{tabular}
\end{spacing}
\caption{Energy gaps to the first excited state for the spin-$1/2$ XXZ model \eqref{eq:H_XXZ} and the Bose-Hubbard model \eqref{eq:H_BH} with $\mu=0.5$ and $U=1$, determined by DMRG.}
\end{table}

Figure~\ref{fig:XXZ-static} displays the accuracies of both methods for fixed values of the total computation cost. The columns refer to the three values of the anisotropy parameter $\Delta$. The top panels present the absolute value of the quasi-exact correlation function $\bra \hS^{-}_{i}\hS^{+}_{0} \ket_\beta$. The panels below display the errors of the purification simulations with $\epsP = 10^{-10}$ and $10^{-12}$ and the errors of several METTS simulations with different truncation thresholds $\epsM$. All curves that appear within a panel and refer to the same temperature are based on simulations of equal total computation costs. For each panel, the METTS truncation thresholds $\epsM$ are chosen such that, for the largest $\epsM$, the truncation error dominates; for the lowest, the statistical error dominates; and for the intermediate $\epsM$, statistical and truncation errors are balanced in an optimal way such that the error is (approximately) minimized for the given computation cost.

For almost all parameters, the matrix product purification simulations yield more accurate results than the best METTS computations. For $\beta=4$, the errors of the methods differ by up to a few orders of magnitude. When lowering the temperature, entanglement and correlation lengths increase. The increased absolute value of the correlation function is reflected in a correspondingly larger absolute error for the simulations based on purifications. The efficiency of the METTS sampling can increase when lowering the temperatures, especially when it gets sufficiently below the energy gap $\Delta E$. In this case, the dimension of the relevant state space to be sampled by METTS is strongly reduced and hence is the statistical error. This is confirmed for the lower temperature $\beta=16$ in Fig.~\ref{fig:XXZ-static}. For the critical (gapless) systems, the purification approach is still more accurate. For the gapped system ($\Delta=3$), the temperature is with $\beta=16$ already substantially below the energy gap, $\beta\Delta E_{\Delta=3}\sim 10$ (cf.\ Table~\ref{tab:gaps}), and METTS can in this case indeed outperform the matrix product purification.

In the introduction, we have already explained why one should expect purifications to be more efficient than METTS at higher temperatures. That METTS can become more efficient at low temperatures is most obvious for the limit $\beta\to \infty$. In this case, every METTS $|\phi_\vn\ket$ for which the initial state $|\vn\ket$ has nonzero overlap with a ground state will simply be this ground state, $|\phi_\vn\ket=|\gs\ket$ (up to truncation errors). The purification, on the other hand, evolves to the purification $|\gs\ket\otimes|\gs\ket_\aux$ of the ground state density matrix $|\gs\ket\bra \gs|$. As the tensor product of two MPS of bond dimension $D$ is an MPS with bond dimension $D^2$, for the same accuracy, $\beta\to\infty$ computation costs for the METTS are reduced by roughly a factor of $D^3$ \footnote{Computation costs for DMRG simulations with open boundary conditions scale as $D^3$ with the bond dimension $D$.}.

The error of the purification changes significantly as a function of the distance $i$. Here, the deviation from the (quasi)-exact reference data is mainly due to the truncations. Generally, when we evaluate a correlator based on a truncated matrix product state, the error can grow at short distances, reaching an $\epsP$-dependent maximum, before starting to decay as the absolute value of the correlator itself becomes very small. On the other hand, METTS simulation errors often remain constant at large distances. Even for the exponentially decaying correlation functions, the METTS error stays well above zero. This is clearly a signature of the statistical error. Cases where the METTS error decays with distance, are usually situations where the truncation error dominates over the statistical error. For the computation costs chosen in our study, this is seen for temperatures well below the energy gap, but one can also observe this behavior when choosing very large truncation thresholds.

Finally, let us shortly discuss an ergodicity issue in the gapped N\'{e}el phase. For $\beta=16$, the system is essentially in its ground state. Hence, the weight of the two degenerate N\'{e}el states in the thermal state becomes significant. The purified state still obeys the $Z_2$ symmetry. Every METTS sample is essentially some linear combination of the ground states. Depending on the choice of the collapse basis $\{|\vn\ket\}$, severe ergodicity problems can occur in the METTS algorithm. As can be shown, even for the rotation-symmetric random bases that we use (described in appendix~\ref{Appx:METTS_collapse}) and which seem to have very good ergodicity properties at first sight, the transition probabilities from one of the degenerate ground states to the other decay exponentially in the system size $L$. Several remedies are available. One can for example avoid this issue by symmetrizing every METTS before measurements. However, in the case of response functions (see section~\ref{Sec:Response}), this would require an additional real-time evolution and, hence, roughly a doubling of the costs. Alternatively, one could, e.g., use the $\hS^x$ and the $\hS^y$ eigenbases for the METTS collapse. Here, we decided to keep the random collapse bases and implicitly average over the two N\'{e}el states by using symmetrized observables; for $\bra \hS^{-}_{i}(t)\hS^{+}_{0}(0) \ket_\beta$, according to
\begin{equation*}
 \bra \hS^{-}_{i}(t)\hS^{+}_{0}(0) \ket_\beta + \bra \hS^{+}_{i}(t)\hS^{-}_{0}(0) \ket_\beta
 = 4 \bra \hS^{x}_{i}(t)\hS^{x}_{0}(0) \ket_\beta,
\end{equation*}
such that the measurement is not sensitive to the broken symmetry.

\subsection{1D Bose-Hubbard model} \label{Sec:Bose-Hubbard}
\begin{figure*}[t]
\label{fig:Bose-Hubbard}
\includegraphics[width=\textwidth]{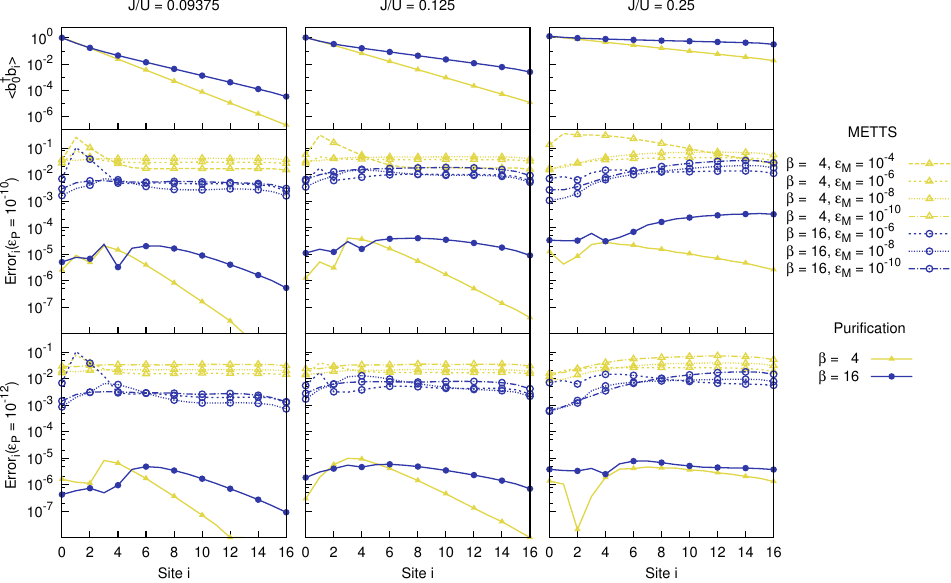}
\caption{Comparing accuracies of METTS and purifications for the static correlation function $\bra \hat{b}^{\dagger}_{0} \hat{b}_{i} \ket_\beta$ in the 1D Bose-Hubbard model \eqref{eq:H_BH} of length $L=32$ at inverse temperatures $\beta=4$, $16$ with $\mu=0.5$ and $U=1$. The local Hilbert spaces were truncated to a maximum of five bosons per site, and site $i=0$ is at the center of the chain. The columns refer to the Mott-insulating phase ($J/U = 0.09375$, left) with one boson per site in the ground state, the phase boundary ($J/U = 0.125$, center) and the superfluid phase ($J/U = 0.25$, right). The corresponding METTS sample set sizes and bond dimensions are specified in Table~\ref{tab:NumberMETTS-BH}.}
\end{figure*}
The Hamiltonian of the one-dimensional (1D) Bose-Hubbard model \cite{Kuehner1998-58,Jaksch1998-81} with open boundary conditions is given by
\begin{equation}\label{eq:H_BH}
\hH = \sum_{i} \left\{-J  \left( \hat{b}^{\dagger}_{i} \hat{b}_{i+1} + \mathrm{h.c.} \right) + \frac{U}{2} \hat{n}_{i}(\hat{n}_{i}-1) - \mu \hat{n}_{i}\right\}
\end{equation}
with ladder operators $\hat{b}_{i}$ obeying $[\hat{b}_i^\pdag,\hat{b}_j^\dag]=\delta_{ij}$ and the number operators $\hat{n}_{i} = \hat{b}^{\dagger}_{i}\hat{b}_{i}$, hopping $J$, onsite interaction $U$, and chemical potential $\mu$.

We set $\mu=0.5$ and $U=1$, and consider three values of the hopping parameter. $J=0.09375$ corresponds to the Mott-insulating phase with one particle per site, $J=0.125$ places the system close to the phase boundary between the Mott-insulating and the superfluid phases, and $J=0.25$ lies in the superfluid phase. In Fig.~\ref{fig:Bose-Hubbard}, we show the comparison between METTS and purifications for the Bose-Hubbard model at $\beta=4$ and $\beta=16$. Again, all curves in a given panel that refer to the same temperature share the same computation cost. In general, the results are similar to those for the XXZ model. However, the differences between the METTS and purifications errors are larger in the comparison for the Bose-Hubbard model. This can be explained with the smaller energy gaps, compared to those in the spin-$1/2$ XXZ model. See Table~\ref{tab:gaps}. For the Bose-Hubbard model at the specified points in the phase diagram, errors of the METTS simulations exceed those of purification simulations by two to three orders of magnitude.

\section{Algorithms for response functions} \label{Sec:Response}
The matrix product purification and METTS methods, as described in sections~\ref{Sec:Purification} and \ref{Sec:METTSalgo}, can be extended to the computation of thermal response functions
\begin{equation}\label{eq:responseFct}
	\bra \hX(t)\hY\ket_\beta\equiv \frac{1}{Z_\beta}\Tr(\dm_\beta \hX(t)\hY)
\end{equation}
with $\hX(t)\equiv \mre^{\mri\hH t}\hX \mre^{-\mri\hH t}$.
For purifications, several schemes have been suggested and analyzed \cite{Barthel2009-79b,Karrasch2012-108,Barthel2012_12,Barthel2013-15}. They have different properties concerning the dependence of computation costs on $\beta$ and $t$. According to the naming introduced in Refs.~\cite{Barthel2012_12,Barthel2013-15}, we address below computation schemes A, B, and C, introduce corresponding schemes for the METTS framework, and discuss their properties.

In Refs.~\cite{Barthel2012_12,Barthel2013-15}, further optimizable classes of schemes have been studied. In comparison to the near-optimal scheme C, they allow to substantially reduce computation costs, for example, for systems with separated energy scales. As there is probably no useful adaptation of them for METTS, they will not be addressed here. In related work, Pi\v{z}orn \emph{et al.}\ \cite{Pizorn2014-16} recently discussed the option of working, for matrix product purifications (equivalently, one can think in terms of matrix product operators \cite{Barthel2013-15}), in the Heisenberg picture, i.e., to compute the vectorization $|X^\dag(t)\ket$ of the evolved operator $\hX^\dag(t)$ and the vectorization $|\dmp_{\beta}\ket$ of the thermal state $\dm_\beta$ to then obtain $\bra \hX(t)\hY\ket_\beta=\bra X^\dag(t)|\hY|\dmp_{\beta}\ket/Z_\beta$. One may notice that this is just a special case of the class of optimizable schemes suggested in Eq.~(9) of Ref.~\cite{Barthel2012_12} or Eq.~(18) of Ref.~\cite{Barthel2013-15}. We will not discuss it further because this Heisenberg picture scheme is in general computationally suboptimal and has no direct METTS equivalent. Specifically, the cost analysis in Refs.~\cite{Barthel2012_12,Barthel2013-15} suggests that the Heisenberg picture scheme will typically reach half the maximum times that can be reached by the optimized schemes or the near-optimal scheme C to be described below.

\subsection{The simple scheme A}
Starting from the matrix product purification $|\dmp_{\beta/2}\ket$ [Eq.~\eqref{eq:purify2}] of the density matrix $\dm_\beta$, according to evaluation scheme A \cite{Barthel2009-79b} for the response function \eqref{eq:responseFct}, we first compute matrix product representations of $\mre^{-\mri\hH t}|\dmp_\beta\ket$ and $\mre^{-\mri\hH t}\hY|\dmp_\beta\ket$ using tDMRG. The response function is then given by the matrix element
\begin{equation}\label{eq:schemeA-MPS}
	\bra \hX(t)\hY \ket_\beta = \frac{1}{Z_\beta}\big[ \bra \dmp_{\beta/2}|\mre^{\mri\hH t} \big] \hX \big[\mre^{-\mri\hH t}\hY|\dmp_{\beta/2}\ket \big].
\end{equation}
The vectorization \eqref{eq:vectorize} of operators corresponds to the isomorphism between the space $\mc{B}(\mc{H})$ of linear maps on the physical Hilbert space $\mc{H}$ and the tensor product space $\mc{H}\otimes \mc{H}_\aux$. It allows us to formulate equivalently (and more intuitively) scheme A in terms of matrix product operators (MPOs). Indicating MPOs with square brackets, scheme A reads in this representation simply
\begin{equation}\label{eq:schemeA-MPO}
	\bra \hX(t)\hY \ket_\beta = \frac{1}{Z_\beta}\Tr\left( \big[ \dm_{\beta/2} \mre^{\mri\hH t}\big] \hX \big[\mre^{-\mri\hH t}\hY\dm_{\beta/2}\big]\right).
\end{equation}
As done in Eqs.~\eqref{eq:schemeA-MPS} and \eqref{eq:schemeA-MPO}, the evolved MPS or MPOs that are used for the evaluation of the response function are always indicated by square brackets in the following.

\begin{figure}[t]
\center
\label{fig:SchemeResponse}
\includegraphics[width = 0.9\columnwidth]{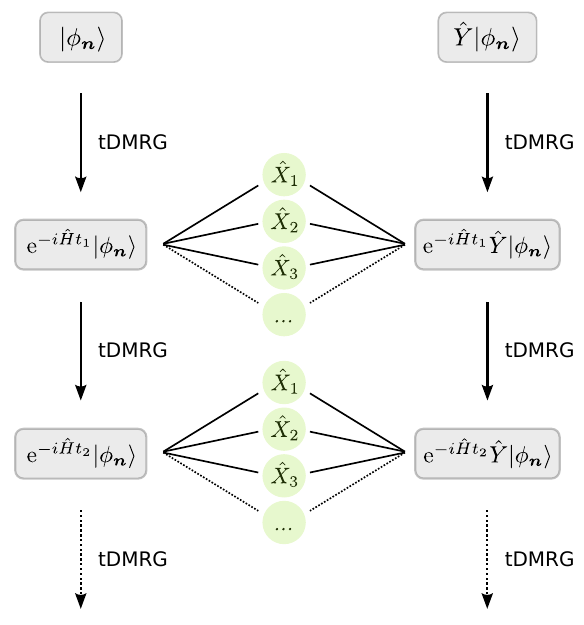}
\caption{The METTS scheme for the evaluation of response functions used in this work (scheme A). We carry out two independent tDMRG-simulations with the METTS sample $|\phi_{\vn}\ket$ and the state $\hY |\phi_{\vn}\ket$ as initial MPS. At intermediate times $t_1, t_2, \dotsc$, we insert operators $\hX_i$ to evaluate. The values obtained need to be averaged over all METTS samples.} 
\end{figure}

The METTS equivalent of scheme A is illustrated in Fig.~\ref{fig:SchemeResponse}. For every sample $|\phi_\vn\ket$, we compute 
\begin{equation} \label{eq:schemeA-METTS}
	\big[\bra \phi_\vn|\mre^{\mri\hH t}\big]\hX\big[\mre^{-\mri\hH t}\hY|\phi_\vn\ket \big]
\end{equation}
using real-time evolution and then average over the values obtained for each sample. In practice, this means that we have to carry out two independent tDMRG simulations, using $|\phi_\vn\ket$ and $\hY|\phi_\vn\ket$ as initial states, up to some maximum time. At any intermediate time-point that we are interested in, we can evaluate the response function $\bra \hX(t)\hY \ket_\beta$ -- even for a set of operators $\{\hX_i\}$ if we wish.

\subsection{Scheme B is not useful for METTS}
\begin{figure}[t]
\center
\label{fig:SchemesCost}
\includegraphics[width = \columnwidth]{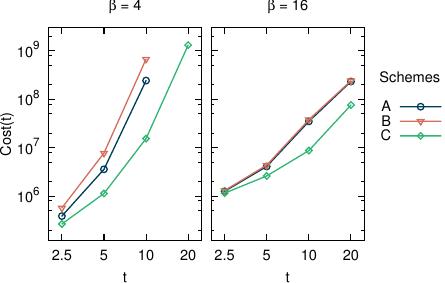}
\caption{Computation costs for different evaluation schemes in the METTS algorithm. For the isotropic Heisenberg antiferromagnet ($\Delta=1$) at inverse temperatures $\beta=4$ (left) and $16$ (right), we show the average cost per sample of the real-time evolution for the evaluation of $\bra \hat{S}^{-}_{0}(t) \hat{S}^{+}_{0}(0) \ket$ in schemes A, B and C (system size $L=32$, truncation threshold $\epsilon_{\mathrm{M}}=10^{-10}$).}
\end{figure}
In the context of matrix product purifications, we have the alternative scheme B \cite{Karrasch2012-108} which reads in the MPO representation
\begin{equation}\label{eq:schemeB-MPO}
	\bra \hX(t)\hY \ket_\beta = \frac{1}{Z_\beta}\Tr\left( \big[ \dm_{\beta/2}\big] \hX \big[\mre^{-\mri\hH t}\hY \mre^{\mri\hH t}\dm_{\beta/2}\big]\right).
\end{equation}
In comparison to scheme A \eqref{eq:schemeA-MPO}, it corresponds to shifting $\exp(\mri\hH t)$ from the first to the second MPO. In Ref.~\cite{Barthel2013-15}, it was explained why this scheme has some advantages at higher temperatures and disadvantages at lower temperatures. The METTS equivalent would be to compute
\begin{equation} \label{eq:schemeB-METTS}
	\big[\bra \phi_\vn|\big]\hX\big[\mre^{-\mri\hH t}\hY \mre^{\mri\hH t}|\phi_\vn\ket \big].
\end{equation}
As exemplified in Fig.~\ref{fig:SchemesCost} and explained in the following, its computation costs are unfortunately strictly higher than those of scheme A \eqref{eq:schemeA-METTS}. Both schemes share the cost for computing $[\mre^{\mri\hH t}|\phi_\vn\ket]$. In scheme A, the evolution of $[\mre^{-\mri\hH t'}\hY|\phi_\vn\ket]$ for $t'=0\to t$ has usually about the same cost, and the entanglement in both states increases typically linearly with $t'$ everywhere in the system, reaching some value $\mc{E}_t$. In scheme B \eqref{eq:schemeB-METTS}, the required evolution of $[\mre^{-\mri\hH t'}\hY \mre^{\mri\hH t}|\phi_\vn\ket]$ for $t'=0\to t$ starts with the high entanglement value $\mc{E}_t$. Due to quasi-locality \cite{Nachtergaele2007-12a,Barthel2012-108b}, the entanglement will then reduce in regions of the lattice that are at sufficient distance from the spatial support of operator $\hY$. It will however remain high or even increase in the vicinity of $\hY$.
\begin{figure*}[t]
\label{fig:XXZ-t}
\includegraphics[width=\textwidth]{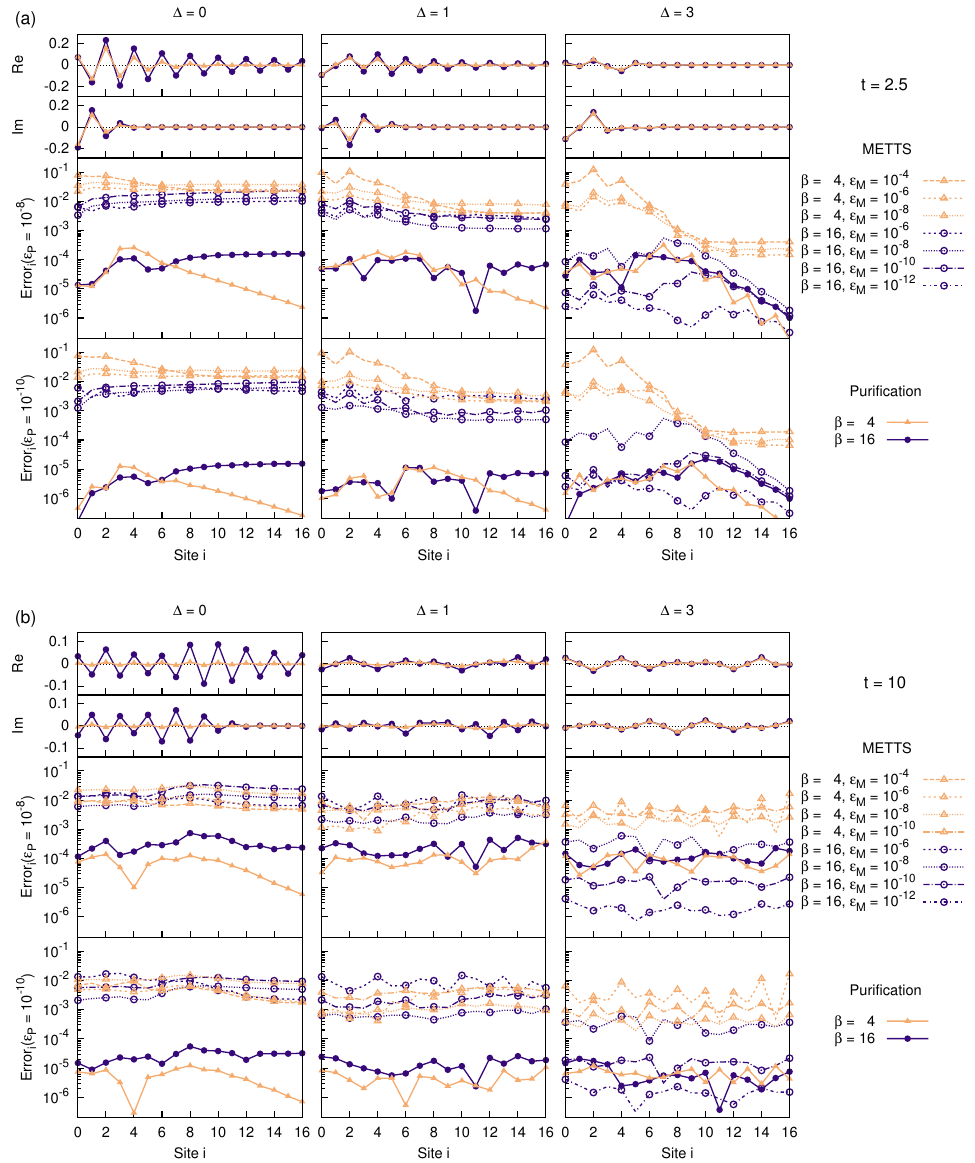}
\caption{Comparing METTS and purifications for the finite-temperature response function $\bra \hS^{-}_{i}(t)\hS^{+}_{0}(0) \ket_\beta$ in the spin-$1/2$ XXZ chain at (a) $t=2.5$ and (b) $t=10$. In the top panels, we show the (quasi-)exact real and imaginary part of the response function. Below we compare the errors as in Fig.~\ref{fig:XXZ-static}. The system size was set to $L=64$ and evaluation scheme A was used.}
\end{figure*}

\subsection{Reaching longer times with scheme C}
Among the classes of optimizable evaluation schemes studied in Refs.~\cite{Barthel2012_12,Barthel2013-15}, the
near-optimal scheme C was found to be very useful and reaches about twice the maximum times reachable with schemes A and B. In the MPO representation, it reads
\begin{multline}\label{eq:schemeC-MPO}
	\bra \hX(t)\hY \ket_\beta = \frac{1}{Z_\beta}\Tr\Big( \big[ \dm_{\beta/2} \mre^{\mri\hH t/2} \hX \mre^{-\mri\hH t/2}\big] \\
	\times\big[\mre^{-\mri\hH t/2}\hY \mre^{\mri\hH t/2}\dm_{\beta/2}\big]\Big).
\end{multline}
In a similar way, we can compute the response function with METTS, by averaging over
\begin{equation} \label{eq:schemeC-METTS}
	\big[\bra \phi_\vn|\mre^{\mri\hH t/2}\hX \mre^{-\mri\hH t/2}\big]\,\big[\mre^{-\mri\hH t/2}\hY \mre^{\mri\hH t/2}|\phi_\vn\ket \big],
\end{equation}
which makes it possible to reach, for the same computational resources, times approximately twice as big as in scheme A.
\begin{figure*}[t]
\label{fig:time-dependence}
\includegraphics[width=\textwidth]{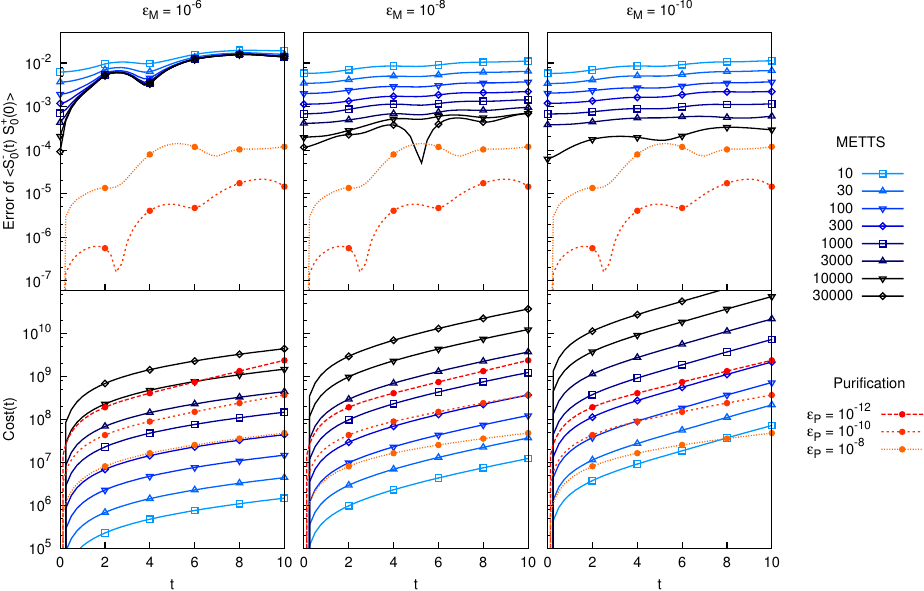}
\caption{Time-dependence of errors and computation costs for the spin-$1/2$ XXZ chain of length $L=32$ at $\Delta=0$ and $\beta=16$. The columns refer to different truncation thresholds in the METTS algorithm. The top panels show the errors for the calculation of $\bra \hS^{-}_{0}(t) \hS^{+}_{0}(0) \ket_\beta$ based on different sample numbers in the METTS algorithm and different truncation thresholds $\epsP$ in the purification. In the lower panels, the corresponding computation costs are presented.}
\end{figure*}

\subsection{Comparison of schemes A, B, and C}
Due to the cyclic property of the trace and the (trivial) fact that the evolution operators commute with density matrices $\dm_\beta$, the results of all three schemes converge up to differences in truncation and Trotter errors (see appendix~\ref{Appx:Truncations}) to the same thermal average \eqref{eq:responseFct}. As the truncations in the real-time evolutions are well controlled, the errors of the simulations based on different schemes do not differ significantly if all other parameters are left unchanged. Statistical errors can however differ somewhat \footnote{For operators $\hY=\hX^\dag$ and Hamiltonians which are real in the chosen onsite basis, the distribution of individual METTS measurement results in schemes A and B will be the same, as the observable $\hX(t) \hX^\dag(0)$, evaluated in scheme A, is then just the transpose of  $\hX(0) \hX^\dag(-t)$, evaluated in scheme B. In general, the distribution of individual measurement results and hence the statistical errors of schemes A, B, and C can differ.}. Figure~\ref{fig:SchemesCost} compares the computation costs of the different evaluation schemes in the METTS algorithm for an autocorrelation function in the isotropic Heisenberg antiferromagnet [$\Delta=1$ in Eq.~\eqref{eq:H_XXZ}]. While scheme B proves to be the least efficient for the parameters considered here, scheme C indeed reduces the computation costs significantly and thus allows for the evaluation of longer maximum times with METTS. 

Nevertheless, we choose to use scheme A in this work because it allows to evaluate response functions $\bra \hX_i(t)\hY \ket_\beta$ for a whole set of operators $\{\hX_i\}$ and all times $t$ (up to the maximum reachable time) with only two tDMRG runs per METTS sample. This is very useful for the analysis of response functions like $\bra \hS^{-}_{i}(t)\hS^{+}_{0}(0) \ket_\beta$, studied in section~\ref{Sec:ResponseXXZ}. In contrast, scheme C requires, for every METTS sample, separate tDMRG runs for every required time $t$ and operator $\hX_i$. This is also a drawback in comparison to scheme C for purifications [Eq.~\eqref{eq:schemeC-MPO}], where one obtains results for all times $t$ with only two tDMRG runs.

\section{Response in the spin-\texorpdfstring{$1/2$}{1\textfractionsolidus 2} XXZ chain} \label{Sec:ResponseXXZ}
Figure~\ref{fig:XXZ-t} compares the accuracies of the response function $\bra \hS^{-}_{i}(t) \hS^{+}_{0}(0) \ket_\beta$ in the XXZ chain as computed via METTS and purification simulations for times $t=2.5$ and $10$. The structure of the plots is as in Fig.~\ref{fig:XXZ-static}, i.e., the columns again refer to the three values of the anisotropy parameter $\Delta=0,1,3$ already considered in section~\ref{Sec:XXZ} and all curves that appear within a panel and refer to the same temperature are based on simulations with the same total computation costs. As the thermal response functions are complex-valued, we show both the real and the imaginary part of the (quasi-)exact results. 

Generally, the error curves closely resemble the results we obtained for the computation of static correlators in Fig.~\ref{fig:XXZ-static}. The purifications are more efficient than METTS except for the gapped system ($\Delta=3$) at the low temperature $\beta=16$.

Figure~\ref{fig:time-dependence} displays the errors and computation costs of both methods as a function of time $t$ in the evaluation of response functions for $\Delta=0$ at inverse temperature $\beta=16$. Each column refers to a truncation threshold $\epsM$ for the METTS algorithm and shows the errors (top) and computation costs (bottom) for different numbers of samples. The purification errors and costs of three different truncation thresholds $\epsP$ are shown in each column.

The convergence of the METTS errors is clearly reminding of the behavior shown in Fig.~\ref{fig:convergence}. As a function of the number of samples, the error is approximately proportional to $1/\sqrt{N}$, until it saturates to an $\epsM$-dependent threshold that can be lowered by decreasing the truncation threshold.

The exponentially growing computation costs in both the METTS approach as well as the matrix product purification approach limit the reachable maximum times, i.e., at least qualitatively, METTS seem to have no favorable properties in this respect.

\section{Conclusions and discussion} \label{Sec:Conclusion}
We have studied properties of the METTS algorithm for the simulation of strongly correlated systems at finite temperatures. There is an interplay of statistical and DMRG truncation errors that one should take into account for efficient simulations. While the optimal truncation threshold $\epsM$ for METTS depends on the specific system studied and the observable that is evaluated, it generally shifts towards lower values of $\epsM$ with increasing total computation cost. As demonstrated, one can also exploit self-averaging of (approximately) translation invariant systems to reduce statistical errors in METTS simulations for static observables.

We have presented a simple scheme for the evaluation of response functions using METTS and two more elaborate schemes, one of which gives access to longer maximum times but needs a separate simulation for every required point in time.

For spin-$1/2$ XXZ chains and the 1D Bose-Hubbard model, we have compared the accuracies and computation costs of the METTS and purification approaches for the evaluation of finite-temperature correlation and response functions. For almost all cases considered here, we found in contrast to indications and expectations expressed in the earlier literature that, for the same total computation cost, the purification approach yields more accurate results than METTS -- often by orders of magnitude. METTS become more efficient only for temperatures well below the energy gap of the system ($\beta\Delta E\gtrsim 10$ in our case). For both methods, we have discussed the temperature dependence of the accuracies, and for correlators also their distance dependence.

It would be interesting to investigate further whether other DMRG approaches for the low-temperature regime, such as computations based on the ground state and a few excited states or a sampling that is restricted to the complement of the ground state space, could outperform the METTS in these cases.

Lastly, please note that also the evolution under time-dependent Hamiltonians, starting from thermal initial states, can be simulated with both, purifications and METTS, in complete analogy to the case of pure states. To this purpose, one can for example employ a Magnus expansion of the time-evolution operator \cite{Alvermann2011-230}.

\addtocontents{toc}{\SkipTocEntry}
\begin{acknowledgments}
We gratefully acknowledge discussions with G.\ Roux, U.\ Schollw\"ock, S.~R.\ White, and M.\ Zvonarev.
\end{acknowledgments}

\appendix

\section{Truncations in our tDMRG simulations} \label{Appx:Truncations}
The results for the observables and computation costs presented here are essentially independent of the chosen time-evolution algorithm. We use time-dependent DMRG (tDMRG) \cite{Vidal2003-10,White2004,Daley2004} and employ a Trotter-Suzuki decomposition \cite{Trotter1959,Suzuki1976} of fourth order, i.e., approximate the time-evolution operator for a (real or imaginary) time step $\Delta\tau$ according to
\begin{equation}
	\mre^{-\Delta\tau\hH}=\prod_k \mre^{-a_k\Delta\tau\hH_\even}\mre^{-b_k\Delta\tau\hH_\odd}+\mc{O}(\Delta\tau^5)
\end{equation}
with suitable coefficients $a_k$, and $b_k$. In this expression, $\hH_\even$ and $\hH_\odd$ contain all Hamiltonian terms on odd and even bonds of the lattice, respectively, such that $\hH=\hH_\even+\hH_\odd$. As all bond terms contained in $\hH_\even$ are mutually commuting, it is simple to apply the unitaries $\mre^{-a_k\Delta\tau\hH_\even}$ (analogously for $\hH_\odd$) to matrix product states (MPS), which have the form
\begin{equation}\label{eq:MPS}
	|\psi\ket = \sum_{\{\sigma_i\}} \Big(\prod_i A^{\sigma_i}_i\Big) |\vec{\sigma}\ket.
\end{equation}
Here, $i$ runs over all lattice sites, and $A^{\sigma_i}_i$ is a $D_i\times D_{i+1}$ matrix. The dimensions $D_i$ are the so-called bond dimensions and the bond dimensions for the left and right ends of the chain are $1$ such that the matrix product $(\prod_i A^{\sigma_i}_i)$ evaluates to a scalar.

When a time-step evolution operator such as $\exp(-a_k\Delta\tau\hH_\even)$ is applied to the MPS, the bond dimensions $D_i$ increase to $\tilde{D}_i\geq D_i$ -- partly for technical reasons, because the resulting state needs to be brought to MPS form, and partly because the entanglement in the state can grow. States with higher entanglement generally require larger bond dimensions. Hence, it is necessary to compress the matrix product representation, i.e., to make a controlled approximation such that the bond dimensions are again reduced to some extent. This can be achieved through a Schmidt decomposition $|\psi\ket=\sum_{j=1}^{\tilde D}\lambda_j|j\ket_L\otimes|j\ket_R$ of the state \cite{Nielsen2000}, which boils down to doing singular value decompositions of the tensors $A_i$. The corresponding reduced density matrices for the left and right parts of the system are $\sum_j\lambda_j^2|j\ket_L\bra j|_L$ and $\sum_j\lambda_j^2|j\ket_R\bra j|_R$, respectively. The bond dimension for the given bipartition of the lattice is then reduced from $\tilde D$ to some value $D'<\tilde D$ by retaining only the $D'$ largest Schmidt coefficients $\lambda_j$ and truncating all smaller ones.
\begin{equation}
	 |\psi_{\operatorname{trunc}}\ket=\sum_{j=1}^{D'}\lambda_j|j\ket_L\otimes|j\ket_R
\end{equation}
The resulting two-norm error is
\begin{equation}
	\left(\frac{\normS{\psi_{\operatorname{trunc}} - \psi}}{\normS{\psi}}\right)^2 = \frac{\sum_{j>D'}\lambda_j^2}{\sum_j\lambda_j^2}.
\end{equation}

In our simulations, we control errors due to truncations by discarding in every step of the algorithm only components with density matrix eigenvalues $\lambda_j^2$ below a predefined truncation threshold. In this way, the bond dimensions become functions of (inverse) temperature and time, $D_i=D_i(\beta,t)$. For purifications, the truncation threshold is denoted by $\epsP$ and by $\epsM$ for the METTS. The employed values are specified in the results sections and corresponding figures. For all imaginary- and real-time evolutions of purifications, we choose time steps of size $\Delta\tau_\mathrm{P}=\Delta t_\mathrm{P}=0.125$ and for the METTS they are $\Delta\tau_\mathrm{M}=\Delta t_\mathrm{M}=0.05$.

\section{Choices for the METTS collapse} \label{Appx:METTS_collapse}
In section~\ref{Sec:METTSalgo}, we reviewed the METTS algorithm for a fixed orthonormal basis $\{|\vn\ket\}$, used in the projective measurements to generate a new product state $|\vn'\ket$ from the current METTS state $|\phi_\vn\ket\equiv \mre^{-\beta\hH/2}|\vn\ket/\sqrt{P_\vn}$. In order to ensure ergodicity and reduce autocorrelation times, it is useful to switch between different measurement bases $\{|\vn^{(k)}\ket\}$ during the sampling \cite{Stoudenmire2010-12}. This is possible, because it simply corresponds to using an overcomplete basis in the representation of the trace in Eq.~\eqref{eq:METTStrace}, i.e., to starting instead from the expression
\begin{equation}
	\bra \hat{O} \ket_\beta = \frac{1}{Z_\beta} \sum_k\pi_k\sum_\vn \bra \vn^{(k)}|\mre^{-\beta \hH} \hat{O} |\vn^{(k)}\ket,
\end{equation}
where $\pi_k\geq 0$ are probabilities ($\sum_k\pi_k=1$) according to which we choose basis $\{|\vn^{(k)}\ket\}$ to collapse a METTS. The rest of the derivation remains unchanged except for extending the detailed balance to the choice of the measurement bases.

In our simulations, we found the following procedure to be robust and efficient. Before every collapse of the wavefunction, we randomly determine measurement bases $\{|n_i^{(k)}\ket,\,n_i=1,\dotsc,d\}$ for the site Hilbert spaces $\operatorname{span}\{|\sigma\ket,\,\sigma=1,\dotsc,d\}$ and then choose $|\vn^{(k)}\ket=\bigotimes_i|n^{(k)}_i\ket$. For every site, the local basis states $|n_i^{(k)}\ket$ are obtained by first choosing $|\tilde{n}_i^{(k)}\ket=\sum_{\sigma=1}^d (a_{n,\sigma}+i b_{n,\sigma})|\sigma\ket$ with $a_{n,\sigma}$ and $b_{n,\sigma}$ drawn from the standard normal distribution. Using the Gram-Schmidt process, these $d$ states are subsequently orthonormalized to obtain $\{|n_i^{(k)}\ket\}$.

\section{Quantification of computation costs} \label{Appx:Computation_cost}
\begin{figure}[t]
\label{fig:fit_comp_time}
\includegraphics[width=\columnwidth]{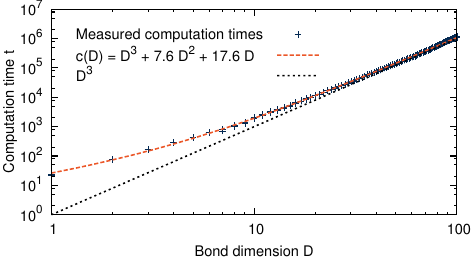}
\caption{Fit to measured computation times $t$ for a single local tDMRG step as a function of the bond dimension $D$. The measured times are normalized such that $\lim_{D\to\infty}t/D^{3} \to 1$. We fit the cost model $c(D)$ of Eq.~\eqref{eq:costModel} and obtain coefficients $a=7.6$, $b=17.6$.}
\end{figure}
The main contribution of this work is to compare the efficiency of the METTS and matrix product purification algorithms. In order to make this comparison as independent as possible from specific details of the implementation and computer architecture, we approximate the computation costs associated with the imaginary- and real-time tDMRG evolutions by integrals over a function of the bond dimensions $D_i=D_i(\beta,t)$, which were described in detail in appendix~\ref{Appx:Truncations}. As the elementary steps in the employed DMRG algorithms are matrix multiplications and singular value decompositions for the MPS matrices $A^{\sigma_i}_i$ [Eq.~\eqref{eq:MPS}], the computation cost scales in leading order with the third power of the bond dimensions. However, small bond dimensions can lead to substantial subleading contributions corresponding to copying matrices etc. In a realistic assessment of the actual computation cost, these need to be taken into account, especially as optimal METTS bond dimensions can be quite small in some cases. Therefore, we model the computation cost $c(D)$, associated with a single local tDMRG step as a function of the bond dimension $D$, by 
\begin{equation}\label{eq:costModel}
	c(D) = D^{3} + a D^{2} + b D.
\end{equation}
As shown in Fig.~\ref{fig:fit_comp_time}, a fit to actual measured computation times yields coefficients $a = 7.6$ and $b=17.6$. While the exact values of $a$ and $b$ depend to some extent on the particular implementation at hand and the employed computer architecture, deviations of the coefficients do not decisively change the results of our study. In essence, the coefficients only depend on the cost scaling of the employed linear algebra routines that are implemented in the highly optimized LAPACK library \cite{LAPACK1999} for the given computer architecture.

\begin{figure}[t]
\label{fig:integrated_comp_times}
\includegraphics[width=0.93\columnwidth]{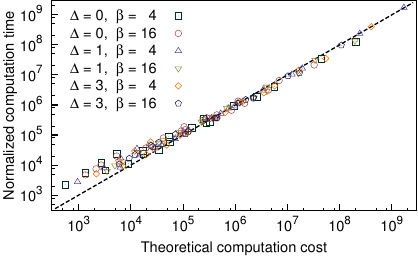}
\caption{Computation time (normalized) versus theoretical computation cost according to Eq.~\eqref{eq:costTheory}. For the XXZ model \eqref{eq:H_XXZ} with $L=32$ sites, we show data points for various truncation thresholds ($\epsM = 10^{-4}$, $10^{-6}$, $10^{-8}$, $10^{-10}$, $10^{-12}$); in each case, we include both the computation of static observables and response functions ($t = 2.5$, $5$, $10$).}
\end{figure}
Based on this, we model the computation costs for an imaginary-time evolution from $\tau=0$ to $\beta/2$ and a subsequent real-time evolution to time $t$, as they occur in the different algorithms (see sections~\ref{Sec:Purification}, \ref{Sec:METTSalgo}, and \ref{Sec:Response}), by
\begin{subequations}\label{eq:costTheory}
\begin{align}
	C_{\operatorname{imag}}(\beta)   &= \int_{0}^{\beta/2}\mathrm{d}\tau \sum_i c(D_{i}(\tau,0))\\
	C_{\operatorname{real}}(\beta,t) &= \int_{0}^{t}\mathrm{d}t' \sum_ic(D_{i}(\beta/2,t')),
\end{align}
\end{subequations}
where $i$ runs over all bonds of the lattice. For schemes that require a certain sequence of imaginary- and real-time evolutions, we sum the corresponding contributions to quantify the total cost. For METTS, we use the average computation cost per METTS and multiply for comparisons by the considered number of METTS.

Figure~\ref{fig:integrated_comp_times} shows that the theoretical computation cost \eqref{eq:costTheory} provides a realistic description of the DMRG computation cost for all simulations, independent of the specific system parameters. Over several orders of magnitude, the agreement is very good; only for extremely short computation times, one obtains slight deviations. This additional overhead is mainly due to the computation time needed for the collapse procedure and the evaluation of observables.

\begin{table*}[p]
\caption{Sample set sizes and bond dimensions for the comparison of static observables in the XXZ chain of length $L=128$ (Fig.~\ref{fig:XXZ-static}). $D_\mathrm{P}$ is the maximum bond dimension of the matrix product purification and $D_\mathrm{M}$ labels the average METTS bond dimension in the center of the chain. $N$ denotes the sample set size that was used in the comparison of the methods.}
\SpaceCaptionTable
\label{tab:NumberMETTS-XXZ}
\centering
\definecolor{darkblue}{RGB}{8,4,84}
\definecolor{darkred}{RGB}{84,19,15}

\newcolumntype{A}{>{\color{darkred}$}r<{$\raggedright}}
\newcolumntype{B}{>{\color{darkblue}$}r<{$\raggedright}}

\setlength{\tabcolsep}{8pt}
\begin{tabular}{| >{$}c<{$}  >{$}c<{$}  >{$}c<{$} || A  B  >{$}r<{$} | A  B  >{$}r<{$} | A  B  >{$}r<{$} |}
\hline
\TT\B \multirow{2}{*}{$\beta$}	& \multirow{2}{*}{$\epsP$}	& \multirow{2}{*}{$\epsM$}	& \multicolumn{3}{ c| }{$\Delta=0$} & \multicolumn{3}{ c| }{$\Delta=1$} & \multicolumn{3}{ c| }{$\Delta=3$} \\
\T\BB 	& 	&  	& D_{\mathrm{P}} & D_{\mathrm{M}} & N &		D_{\mathrm{P}} & D_{\mathrm{M}} & N &	D_{\mathrm{P}} & D_{\mathrm{M}} & N		\\ \hhline{|===#===|===|===|}
\T 4  & 10^{-10} & 10^{-4}  & 22   & 1.0   & 190   & 29   & 1.0  & 352  & 42   & 1.0  & 473  \\
4     & 10^{-10} & 10^{-6}  & 22   & 2.0   & 55    & 29   & 2.1  & 82   & 42   & 2.4  & 188  \\
4     & 10^{-10} & 10^{-8}  & 22   & 3.8   & 23    & 29   & 4.5  & 38   & 42   & 4.8  & 75   \\
4     & 10^{-10} & 10^{-10} & 22   & 5.5   & 13    & 29   & 7.9  & 15   & 42   & 8.0  & 31   \\
4     & 10^{-12} & 10^{-4}  & 28   & 1.0   & 430   & 51   & 1.0  & 1367 & 73   & 1.0  & 1841 \\
4     & 10^{-12} & 10^{-6}  & 28   & 2.0   & 124   & 51   & 2.1  & 319  & 73   & 2.4  & 734  \\
4     & 10^{-12} & 10^{-8}  & 28   & 3.8   & 54    & 51   & 4.5  & 150  & 73   & 4.8  & 292  \\
\B 4  & 10^{-12} & 10^{-10} & 28   & 5.5   & 30    & 51   & 7.9  & 62   & 73   & 8.0  & 121  \\
\hline
\T 16 & 10^{-10} & 10^{-6}  & 61   & 2.0   & 486   & 108  & 2.2  & 1611 & 72   & 2.4  & 926  \\
16    & 10^{-10} & 10^{-8}  & 61   & 5.2   & 143   & 108  & 7.0  & 330  & 72   & 4.7  & 313  \\
16    & 10^{-10} & 10^{-10} & 61   & 10.2  & 44    & 108  & 14.4 & 89   & 72   & 7.7  & 105  \\
16    & 10^{-10} & 10^{-12} & 61   & 15.8  & 16    & 108  & 23.6 & 30   & 72   & 12.2 & 38   \\
16    & 10^{-12} & 10^{-6}  & 110  & 2.0   & 2308  & 204  & 2.2  & 8301 & 126  & 2.4  & 4904 \\
16    & 10^{-12} & 10^{-8}  & 110  & 5.2   & 683   & 204  & 7.0  & 1702 & 126  & 4.7  & 1661 \\
16    & 10^{-12} & 10^{-10} & 110  & 10.2  & 210   & 204  & 14.4 & 462  & 126  & 7.7  & 560  \\
\B 16 & 10^{-12} & 10^{-12} & 110  & 15.8  & 79    & 204  & 23.6 & 155  & 126  & 12.2 & 203  \\
\hline
\end{tabular}
\end{table*}
\begin{table*}[p]
\caption{Sample set sizes and bond dimensions for the comparison of static observables in the Bose-Hubbard model with size $L=32$ (Fig.~\ref{fig:Bose-Hubbard}). $D_\mathrm{P}$, $D_\mathrm{M}$, and $N$ are defined as in Table~\ref{tab:NumberMETTS-XXZ}.}
\SpaceCaptionTable
\label{tab:NumberMETTS-BH}
\centering

\definecolor{darkblue}{RGB}{8,4,84}
\definecolor{darkred}{RGB}{84,19,15}

\newcolumntype{A}{>{\color{darkred}$}r<{$\raggedright}}
\newcolumntype{B}{>{\color{darkblue}$}r<{$\raggedright}}

\setlength{\tabcolsep}{8pt}
\begin{tabular}{| >{$}c<{$}  >{$}c<{$}  >{$}c<{$} || A  B  >{$}r<{$} | A  B  >{$}r<{$} | A  B  >{$}r<{$} |}
\hline
\TT\B \multirow{2}{*}{$\beta$}	& \multirow{2}{*}{$\epsP$}	& \multirow{2}{*}{$\epsM$}	& \multicolumn{3}{ c| }{$J/U = 0.09375$} & \multicolumn{3}{ c| }{$J/U = 0.125$} & \multicolumn{3}{ c| }{$J/U = 0.25$} \\
\T\BB 	& 	&  	& D_{\mathrm{P}} & D_{\mathrm{M}} & N &		D_{\mathrm{P}} & D_{\mathrm{M}} & N &	D_{\mathrm{P}} & D_{\mathrm{M}} & N		\\ \hhline{|===#===|===|===|}
\T 4	& 10^{-10}	& 10^{-4}	& 16 & 1.0  &  166	& 21 & 1.0  &  236	& 44  & 1.0  &  1363	\\
4	& 10^{-10}	& 10^{-6}	& 16 & 2.1  &  52	& 21 & 2.6  &  52	& 44  & 3.0  &  239	\\
4	& 10^{-10}	& 10^{-8}	& 16 & 3.3  &  26	& 21 & 3.7  &  31	& 44  & 5.4  &  92	\\
4	& 10^{-10}	& 10^{-10}	& 16 & 5.5  &  10	& 21 & 6.3  &  12	& 44  & 9.8  &  29	\\
4	& 10^{-12}	& 10^{-4}	& 31 & 1.0  &  588	& 40 & 1.0  &  1128	& 66  & 1.0  &  4684	\\
4	& 10^{-12}	& 10^{-6}	& 31 & 2.1  &  186	& 40 & 2.6  &  251	& 66  & 3.0  &  821	\\
4	& 10^{-12}	& 10^{-8}	& 31 & 3.3  &  93	& 40 & 3.7  &  148	& 66  & 5.4  &  317	\\
\B 4	& 10^{-12}	& 10^{-10}	& 31 & 5.5  &  38	& 40 & 6.3  &  59	& 66  & 9.8  &  102	\\
\hline
\T 16   & 10^{-10}	& 10^{-6}	& 33 & 2.1  &  246	& 44 & 2.7  &  324	& 99  & 3.1  &  2019 	\\
16	& 10^{-10}	& 10^{-8}	& 33 & 3.2  &  119	& 44 & 3.8  &  176	& 99  & 6.5  &  543	\\
16	& 10^{-10}	& 10^{-10}	& 33 & 7.0  &  33	& 44 & 8.5  &  42	& 99  & 11.9 &  142	\\
16	& 10^{-12}	& 10^{-6}	& 57 & 2.1  &  1008	& 81 & 2.7  &  1651	& 187 & 3.1  &  12204 	\\
16	& 10^{-12}	& 10^{-8}	& 57 & 3.2  &  487	& 81 & 3.8  &  898	& 187 & 6.5  &  3285 	\\
\B 16	& 10^{-12}	& 10^{-10}	& 57 & 7.0  &  138	& 81 & 8.5  &  215	& 187 & 11.9 &  864	\\
\hline
\end{tabular}
\end{table*}

\section{Exact versus quasi-exact reference data} \label{Appx:ReferenceData}
\begin{figure}[h]
\label{fig:quasi_exact}
\includegraphics[width = 0.93\columnwidth]{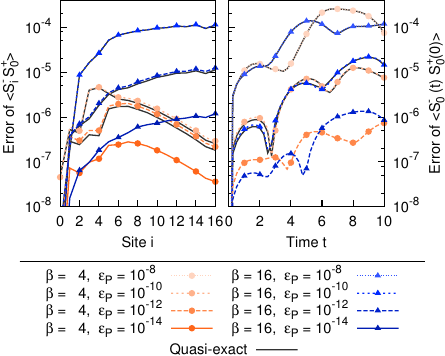}
\caption{Deviations of purification results for different truncation thresholds $\epsP$ from exact as well as quasi-exact reference data in an XX chain [$\Delta = 0$ in Eq.~\eqref{eq:H_XXZ}] of length $L=32$ at inverse temperatures $\beta = 4$ and $16$. The static correlator $\bra \hS^{-}_{i} \hS^{+}_{0} \ket_\beta$ (left) and the response function $\bra \hS^{-}_{0}(t)\hS^{+}_{0}(0)\ket_\beta$ (right) are compared against an exact solution (colored curve with symbols) as well as against quasi-exact purification data (corresponding black solid curves). Due to the high precision of the quasi-exact data, these curves overlap partially.}
\end{figure}

In our assessments of efficiencies, we need to quantify the errors of observables and for this purpose, reliable exact or quasi-exact data are indispensable. For the non-interacting point $\Delta=0$ of the XXZ model \eqref{eq:H_XXZ}, the XX model, we can evaluate all observables exactly. As the system is for this case in a Gaussian state, all correlation functions are determined by the single-particle Green's function due to Wick's theorem \cite{Fetter1971}. Exploiting this, we can obtain static correlations as well as response functions by evaluating Pfaffian determinants of matrices that contain elements of the single-particle Green's function \cite{Caianello1952,Green1964,Stolze1995}. The colored curves in Fig.~\ref{fig:quasi_exact}, marked with triangles and circles, show the errors of matrix-product purification simulations for different truncation thresholds $\epsP$ as compared to such exact results.

Although the XXZ model \eqref{eq:H_XXZ} is Bethe-Ansatz integrable \cite{Bethe1931} for all values of $\Delta$, no exact expressions for longer-ranged static correlators or response functions are available. The same is true for the Bose-Hubbard model \eqref{eq:H_BH}, which is not (known to be) integrable. For these cases, we use very accurate purification simulations to obtain quasi-exact reference data. In particular, we used the truncation threshold $\epsP = 10^{-14}$ during imaginary-time evolutions and $\epsP = 10^{-12}$ during (subsequent) real-time evolutions. ($\epsP$ has been defined in appendix~\ref{Appx:Truncations}.) To justify this procedure, we show in Fig.~\ref{fig:quasi_exact} with black solid lines (no symbols) also the errors of the other simulations compared to this quasi-exact data.
These curves deviate indeed only very slightly from the colored curves that correspond to the comparison to the exact results.

\bibliographystyle{prsty}

\end{document}